# Co-planar spin-polarized light emitting diode

B. Kaestner†*, J. Wunderlich*, Jairo Sinova‡ and T. Jungwirth§#

†*National Physical Laboratory, Teddington T11 0LW, UK*

**Hitachi Cambridge Laboratory, Cambridge CB3 0HE, UK*

‡*Deparment of Physics, Texas A&M University, College Station, TX 77843-4242, USA*

§*Institute of Physics ASCR, Cukrovarnická 10, 162 53, Praha 6, Czech Republic*

#*School of Physics and Astronomy, University of Nottingham, Nottingham NG7 2RD, UK*

**Abstract**

Studies of spin manipulation in semiconductors has benefited from the possibility to grow these materials in high quality on top of optically active III-V systems. The induced electroluminescence in these layered semiconductor heterostructures has been used for a reliable spin detection. In semiconductors with strong spin-orbit interaction, the sensitivity of vertical devices may be insufficient, however, because of the sepration of the spin aligner part and the spin detection region by one or more heterointerfaces and becuse of the short spin coherence length. Here we demostrate that higly sensitive spin detection can be achieved using a lateral arrangement of the spin polarized and optically active regions. Using our co-planar spin-polarized light emitting diodes we detect electrical field induced spin generation in a semiconductor heterojunction two-dimensional hole gas. The polarization results from spin asymmetric recombination of injected electrons with strongly SO coupled two-dimensional holes. The possibility to detect magnetized Co particles deposited on the co-planar diode structure is also demonstrated.

The study of nonequilibrium carrier spin polarization in nonmagnetic semiconductors provides a detailed experimental test of theoretical descriptions of fundamental properties such as the spin-orbit (SO) interaction [1] and may lead to novel device applications [2]. The use of light emitting diodes (LEDs) as detectors of electron-spin polarization has become increasingly popular as it is less ambiguous than schemes based on resistance measurements and allows angle resolved studies [3,4]. The detection of spin-polarization phenomena is done by measuring circular polarization $P_C$. Due to optical selection rules, a finite $P_C$ along a given direction of the propagating light indicates a finite spin-polarization $P_{Spin}$ in this direction of carriers involved in the recombination. Both $P_C$ and $P_{Spin}$ are defined as the ratio of the difference and the sum of the polarization or spin resolved components of the respective quantity.

So far, LEDs applied for spin-detection have been built within vertical devices which are in many cases inpractical for studying spin-related transport phenomena in strongly SO coupled low-dimensional systems [5,6,7,8]. Optical detection via spin-polarized LEDs (spin-LEDs) requires a lateral arrangement of the low-dimensional channel and the optically active region.

Several lateral fabrication schemes have been demostrated to date. Lateral junctions between a 2D electron gas (2DEG) and a 2D hole gas (2DHG) have been realized [9] exploiting the amphoteric nature of Si in GaAs [10]. Similar geometries were produced by focused-ion molecular-beam epitaxy [11]. Checcini *et al.* recently demonstrated a scheme where acceptors in a *p*-type modulation doped AlGaAs – GaAs heterostructure are locally substituted by donors from an ohmic contact [12].

In this paper we demonstrate a device structure where the 2DEG and 2DHG are vertically offset, which tends to improve the confinement of light emission and allows different probing regimes as shown below. The fabrication is based on the previously developed method of selectively etching a *vertical p-i-n* modulation doped heterostructure [13]. The detailed description of the structure and fabrication can be found in [14, 15]. The schematic cross-section is illustrated in Fig. 1(a) which consists of a 90 nm wide intrinsic GaAs region sandwiched between 50nm *p*-type $Al_{0.5}Ga_{0.5}As$ (nominal Be-concentration is $8 \times 10^{18}$ cm$^{-3}$ with a 3nm spacer layer) and *n*-δ-doped $Al_{0.3}Ga_{0.7}As$ (5nm spacer layer). In the as-grown structure the 2DEG is depleted, such that only a 2DHG forms at the upper *i*-GaAs interface. Removing the 50nm *p*-type $Al_{0.5}Ga_{0.5}As$ by wet etching depletes the 2DHG and lowers the conduction band energy so that a 2DEG forms at the lower interface and a quasi-lateral junction forms. The onset of the current is accompanied by electroluminescence (EL) from the unetched region near the junction step edge. Despite the vertical *p-n* doping profile, electroluminescence is confined to the step edge as a result of the local vertical voltage drop varying along the channel. The current in the 20 μm wide channel, $I_{LED} \approx 100$ μA, is dominated by electrons moving from the *n*- to *p*-region mainly because ionized donors exist in addition to acceptors in the unetched region. The advantage of the 90nm vertical off-set between 2DEG and 2DHG is the possibility to scan the spin-polarization of the 2D carrier system underneath the unetched region with injected carriers from the etched region, as we now discuss in detail.

The application of a spin-LED for quantifying the electron spin polarization relies on the correct identification of the spectral components in the electroluminescence (EL) spectrum, which may include free and impurity bound excitons as well as phonons and various impurity levels participating in the recombination [16]. In order to gain information on $P_{Spin}$ from the $P_C$ measurement the quantum selection rules for the corresponding recombination must be known. The EL spectrum of our device was analysed in [17] and is shown in Fig. 1 (b). It consists of 3 regions labelled as *A*, *B* and *C*. Peaks *A* and *B* originate from the recombination of 3D electrons to the spin-split 2D heavy hole subband [18] and their energy depends strongly on temperature and $I_{LED}$ (not shown). The same processes are believed to contribute to Region *C* in addition to a complex of exciton transitions. Peak *A* might also include transitions between the 2DEG to acceptor or donor to 2DHG.

The structure can be operated in two regimes, which ideally can be spectrally separated: injection of spin polarized carriers into the optically active region or probing spin polarization of carriers that are already contained in this region with carriers injected from outside. We now discuss the latter process in which spin polarization of the 2DHG is measured. Microscopic calculations [17] of the 2DHG energy structure is shown in Fig. 2(a) together with the $k_y$-dependent spin-orientation. The spin-splitting is a result of the SO coupling due to asymmetric confinement. Except for a small region near $k_y$=0, spins are oriented in-plane and perpendicular to the *k*-vector. One heavy-hole (HH) and one light-hole (LH) bound state forms in the quantum well and the HH 2D subband is occupied by holes assuming measured hole density, $p_{2D}=2\times10^{12}$ cm$^{-2}$. The *k*-dependent spin orientation becomes apparent when the *p-n* junction is biased. While electron occupation is strongly shifted in the *k*-space, hole distribution is expected to be closer to thermal equilibrium. The probability of radiative recombination (illustrated by vertical lines in Fig. 3(a)) of electrons with holes differs for the two spin orientations of the electrons at a given **k** because the occupation of holes is different between the HH+ and HH- bands. The *k*-region where this effect is most pronounced is marked by the hatched pattern, the shaded area illustrating the electron-state occupation. In this sense the hole-spin is probed within a certain *k*-range and at the same time the spin asymmetric recombiantion leads to a net spin polarization in *x*-direction [20]. The accompanied emission of circularly polarized light from the quantum well serves as a signature of spin polarization. As apparent from the above assignment of the peaks the relevant spectral region is peak *B*, where the influence of impurities can safely be neglected and of which $P_C$ was measured as a function of the detection direction in the plane perpendicular to $I_{LED}$. The measurement was carried out at $T \approx 4$ K and is shown in Fig. 2(b). As expected the circular polarization $P_C$ increases as the detection direction is changed from perpendicular, $\alpha=0$, to in-plane, $\alpha=\pm 90^o$.

We conclude that our co-planar spin-LED can be used to study spin-related phenomena in low-dimensional structures with strong SO coupling. We used the device for detecting electrical field induced spin polarization in a 2DHG. The phenomenology of the observed effect is consistent with theoretical predictions based on microscopic theory of the 2DHG band structure.

In the final paragraphs we demonstrate that co-planar LEDs can be used to study a wide range of spin related phenomena. In particular, we use the diode to detect polarization of a magnetic particle (20nm cobalt film, covered by a 4nm gold layer) deposited on the semiconductor structure. An in-plane magnetic field was applied to magnetize the particle along the field direction. Magnetic force microscopy confirmed that after switching the external field off, the particle remained polarized. The perpendicular component of the generated strayfield at the lower AlGaAs/GaAs interface is estimated to be about 200mT. The emitted light perpendicular to the surface of the forward biased *p-n* junction in zero external field was analysed with respect to $P_C$. The two curves representing the cases of two antiparallel magnetization directions of the particle are displayed in Fig. 3. One of the curves shows a visible effect near the high-energy side of the spectrum (excitonic region) of about 1% whereas the other magnetization orientation does not lead to a measurable effect. This might be due to the spatial variation of the magnetization across the particle for the two different magnetization directions, i.e. the magnetic charge forming in opposite corners of the particle. The large sensitivity of our spin-LED to out-of-plane magnetic fields has also been tested directly[19]. A similar value of $P_C \approx 0.5\%$ in this spectral range was measured in the absence of a magnetic particle under the applied external magnetic field $B$=200mT.

We acknowledge support from the Grant Agency of the Czech Republic under Grant No. 202/05/0575, from Academy of Sciences of the Czech Republic under Institutional Support No. AV0Z10100521, and from the UK EPSRC through Grant GR/S81407/01.

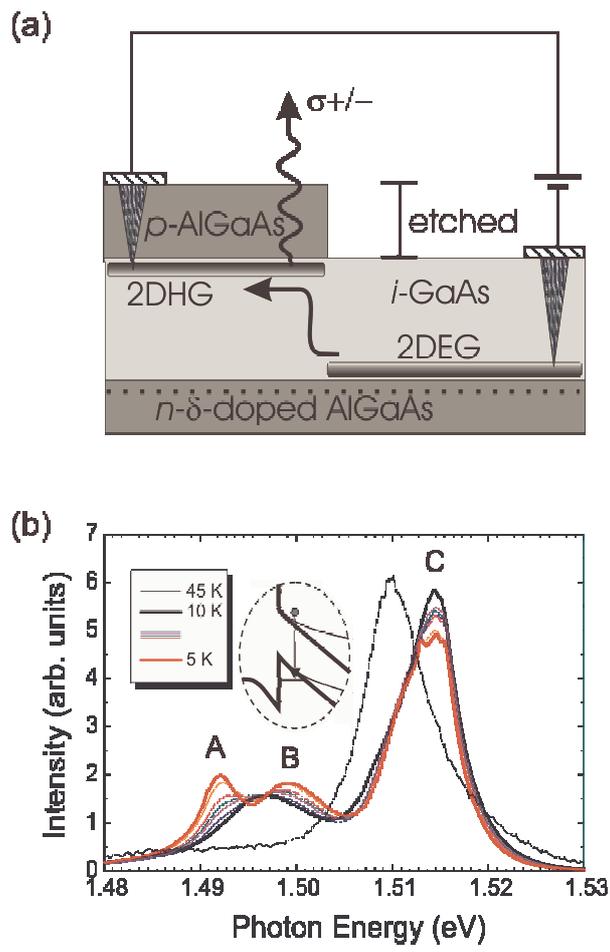

Figure 1

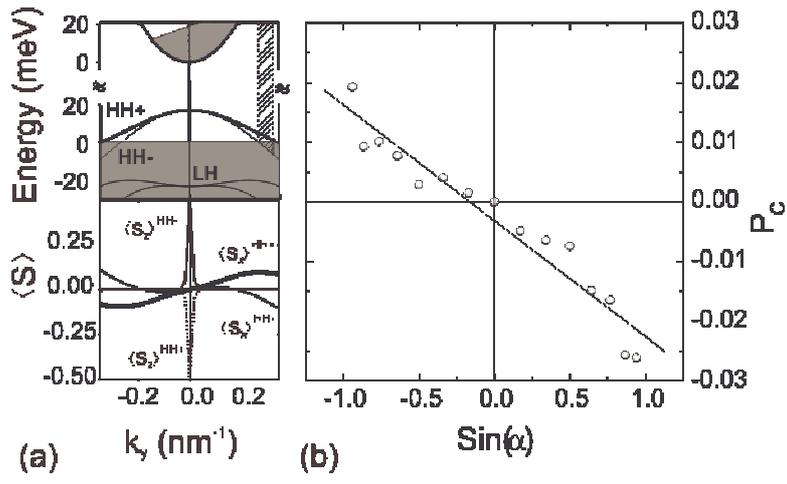

Figure 2

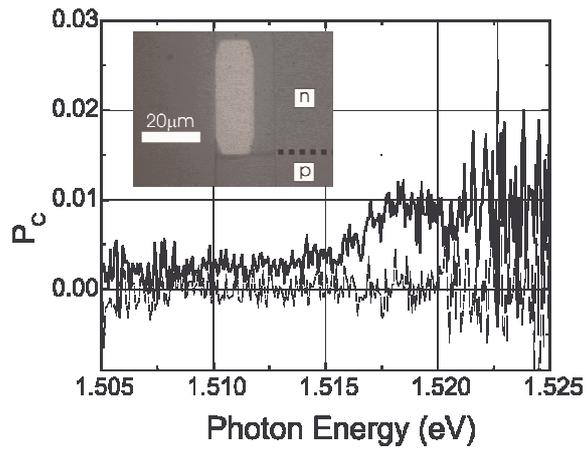

Figure 3

**List of Figures:**

**Figure 1**

Basic characteristics of the LED. (a) The schematic cross-section of the structure, indicating light emission from the *p*-side of the step-edge and electron injection into the unetched region in forward bias. (b) Electroluminescence spectrum of the forward biased junction for different temperatures. The inset shows the 3D electron to 2D hole transition allowing direct probing of the polarization of the 2D holes.

**Figure 2**

Optoelectronic spin-generation and *k*-selective spin-detection. (a) Energy as function of $k_y$ (parallel to $I_{LED}$). Shaded regions illustrate occupation of states under forward bias. Circular polarized light emission is induced by transitions (vertical lines) from the asymmetrically occupied conduction band to the spin-split HH subband. The theoretical spin-polarization components of the HH+ and HH- subbands is shown in below [13]. (An infinitesimal magnetic field along *z*-direction was added to break the degeneracy at **k=0**). (b) Circular polarization of peak *B* as a function of detection angle.

**Figure 3**

Circular polarization of the LED after external inplane magnetic field was switched off for two different ferromagnetic states of the Cobalt particle. The inset shows an optical micrograph of the particle and its position relative to the step-edge separating *p*- and *n*-type regions.